\documentstyle[prl,aps,multicol,epsf]{revtex}
\begin{document}
\draft
\title{Mobility-Dependence of the Critical Density in Two-Dimensional Systems:\\ An
Empirical Relation.}
\author{M.~P.~Sarachik}
\address{Physics Department, City College of the City
University of New York, New York, New York 10031}

\date{\today}
\maketitle

\begin{abstract}
For five different electron and hole systems in two dimensions (Si MOSFET's, p-GaAs,
p-SiGe, n-GaAs and n-AlAs), the critical density, $n_c$ that marks the onset of
strong localization is shown to be a single power-law function of the scattering
rate $1/\tau$ deduced from the maximum mobility.  The resulting curve defines the
boundary separating a localized phase from a phase that exhibits metallic
behavior.  The critical density $n_c \rightarrow 0$ in the limit of infinite
mobility.

\end{abstract}

\pacs{PACS numbers: 71.30.+h, 73.40.Qv}

\begin{multicols}{2}

The unusual properties of dilute two-dimensional (2D) systems of electrons and holes
have been the subject of intense scrutiny and debate \cite{review} since the
discovery by Kravchenko {\it et al} \cite{krav} in 1995 of an apparent
metal-insulator transition in silicon MOSFET's at a critical electron density $n_c
\approx 0.8 \times 10^{11}$ cm$^{-2}$.  The temperature dependence of the
resistivity was found to be metallic ($d\rho/dT>0$) for densities greater than
$n_c$ and localized (($d\rho/dT<0$) below $n_c$.  Similar transitions to
metallic behavior were subsequently reported in a number of other 2D systems,
including p-SiGe \cite{coleridge}, p-GaAs \cite{hanein,simmons}, n-AlAs
\cite{papadakis} and n-GaAs \cite{hanein2}.  This note calls attention to an
empirical relation between the scattering rate deduced from maximum sample mobility
and the density $n_c$ for the onset of strong localization.  For five different
materials, data ranging over three orders of magnitude are shown to lie on a single
curve, suggesting that the onset of strong localization is governed by the same
physics in all two-dimensional systems of electrons and holes studied to date.

The critical density $n_c$ differs for different materials; it is of the order of
$10^{11}$ cm$^{-2}$ in n-silicon and p-SiGe, an order of magnitude lower for
p-GaAs, and even lower for n-GaAs.  It has also been reported \cite{pudalov,jiang}
that the precise value of $n_c$ depends on sample quality, with larger mobility
samples having smaller $n_c$.  On a log-log scale, Fig. 1 shows the critical
density $n_c$ plotted as a function of scattering rate calculated from the measured
peak mobility through the relation $\tau = \mu m^* c/e$ using data reported by
different groups for silicon MOSFET's \cite{pudalov2,okamoto,klapwijk}, p-GaAs
\cite{hanein,simmons,jiang,yoon,mills}, p-SiGe \cite{coleridge}, n-AlAs
\cite{papadakis} and n-GaAs \cite{hanein2,newsimmons}.  

Although there are some systematic differences for samples of the same material
deriving from different sources, the data for five different material systems lie
approximately on a single straight line on a log-log plot, indicating power law
behavior of the form $n_c = A \times (1/\tau)^\beta$ (here $A \approx 2240$
cm$^{-2}$ s$^\beta$ and $\beta=0.67$).  This implies that $n_c
\rightarrow 0$ in the limit of zero scattering rate ({\it i. e.} infinite
mobility).  

\vbox{
\vspace{0.2in}
\hbox{
\hspace{0in} 
\epsfxsize 3.in \epsfbox{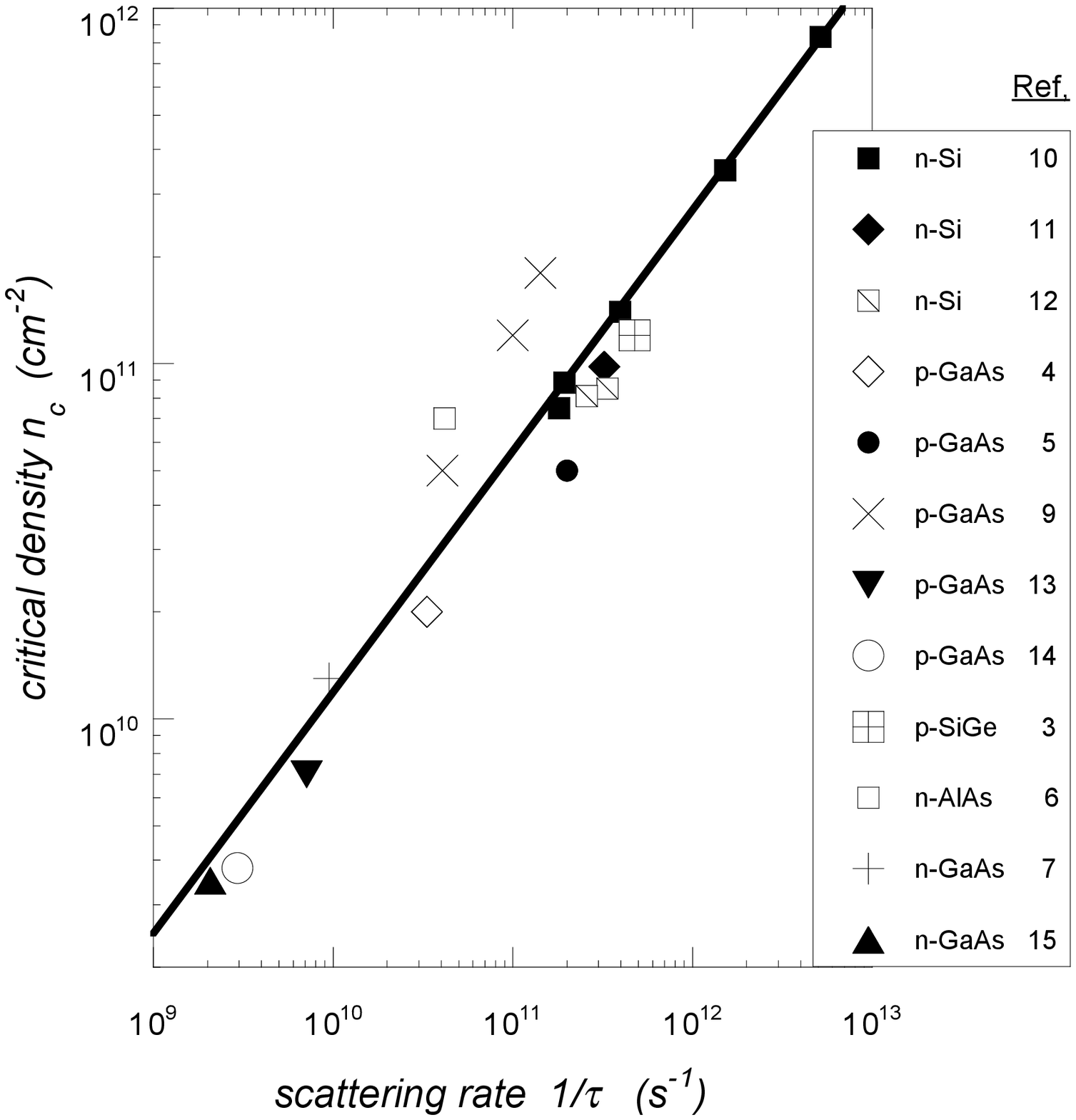} 
}
\vspace{0.1in}
}
\refstepcounter{figure}
\parbox[b]{3.3in}{\baselineskip=12pt FIG.~\thefigure.
Critical density $n_c$ versus scattering rate $1/\tau$ on a double
logarithmic scale.
\vspace{0.10in}
}
\label{1}

\vspace{0in}

The curve of Fig. 1 defines the boundary between a phase where the resistivity
exhibits metallic temperature dependence (above the line) and a strongly
localized phase (below the curve).  The nature of these phases remains unclear. 
Both phases display unusual behavior, including an
exceptionally strong response to in-plane magnetic fields \cite{review} and full
spin polarization above relatively moderate fields in the metallic phase
\cite{okamoto,SdH,tutuc}.

The empirical relation reported in this note implies that a
transition to strong localization is triggered by the degree of disorder in the
system.  This may occur when the Ioffe-Regel condition for localization
($k_F l \approx 1$) is reached, or percolation may play a role, or some
other physical process may enter.   One should not assume, however, that
interactions are unimportant.  It is well known that disorder and interactions are
often inextricably linked: as the transition is approached, the carriers screen
each other less and less effectively, and the interactions between them become
strong.  It should also be noted that the interaction strength is a
"hidden" variable.  The diagram of Fig. 1 allows for values of
the interaction parameter $r_s=E_{int}/E_F$ within the metallic region (above the
line) that are sufficiently large to trigger Wigner crystallization.   Fig. 1
provides a working criterion for recognizing such a transition: it would occur
at a density and for a scattering rate which places the transition above the
line and not on the boundary.

I am grateful to H. W. Jiang, S. V. Kravchenko and S. A. Vitkalov for extensive
discussions and illuminating insights.  This work was supported by grant
DOE-FG02-84-ER45153.  Partial support was also provided by NSF grant DMR-9803440.

\end{multicols}

\end{document}